%% file: main.tex
\documentclass{article}

\usepackage{arxiv}

\usepackage[utf8]{inputenc} 
\usepackage[T1]{fontenc}    
\usepackage{hyperref}       
\usepackage{url}            
\usepackage{booktabs}       
\usepackage{amsfonts}       
\usepackage{nicefrac}       
\usepackage{microtype}      
\usepackage{graphicx}
\usepackage[square, numbers, sort&compress]{natbib}
\usepackage{doi}
\usepackage{amsmath}
\usepackage{amsthm}
\usepackage{listings}
\usepackage[ruled,vlined]{algorithm2e}
\usepackage{multirow}
\usepackage[table]{xcolor}

\theoremstyle{definition}
\newtheorem{definition}{Definition}[section]

\theoremstyle{remark}
\newtheorem{notation}{Notation}[section]

\theoremstyle{remark}
\newtheorem{remark*}{Remark}[section]

\theoremstyle{definition}
\newtheorem{example}{Example}[section]

\title{So Many Ways to Multiply $3\times3$ Matrices: A Large Dataset of Rank-$23$ Schemes}

\author{\href{https://orcid.org/0000-0001-8047-0114}{\includegraphics[scale=0.06]{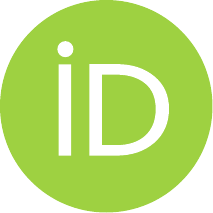}\hspace{1mm}Andrew Perminov}\\
	Research Center for TAI\\
	Institute for System Programming\\
	Moscow \\
	\texttt{perminov@ispras.ru}
}

\renewcommand{\headeright}{}

\newcommand{\heuleSchemes}{17\,376}
\newcommand{\heulePatterns}{387}
\newcommand{\heuleF}{303}
\newcommand{\heuleG}{91}
\newcommand{\heuleMinNnz}{139}
\newcommand{\heuleMaxNnz}{186}

\newcommand{\hellsBellsSchemes}{15\,316}
\newcommand{\hellsBellsPatterns}{568}
\newcommand{\hellsBellsF}{215}
\newcommand{\hellsBellsG}{83}
\newcommand{\hellsBellsMinNnz}{139}
\newcommand{\hellsBellsMaxNnz}{200}

\newcommand{\unionPatterns}{771}

\newcommand{\commonPatterns}{184}

\newcommand{\numSchemes}{62\,029}
\newcommand{\numOrbitsLB}{78\,634}
\newcommand{\countZT}{21\,585}
\newcommand{\countZ}{22\,130}
\newcommand{\countQ}{18\,314}
\newcommand{\percentZT}{34.80}
\newcommand{\percentZ}{35.68}
\newcommand{\percentQ}{29.52}

\newcommand{\numF}{7\,930}
\newcommand{\numFZT}{3\,421}
\newcommand{\numFZ}{5\,072}
\newcommand{\numFQ}{4\,774}
\newcommand{\numG}{530}
\newcommand{\numGZT}{339}
\newcommand{\numGZ}{441}
\newcommand{\numGQ}{466}

\newcommand{\minNnz}{139}
\newcommand{\maxNnz}{278}

\newcommand{\minNnzNonZT}{149}
\newcommand{\modeNnzZT}{175}
\newcommand{\modeNnzZ}{174}
\newcommand{\modeNnzQ}{178}

\newcommand{\rk}{\operatorname{rk}}
\newcommand{\tr}{\operatorname{tr}}
\newcommand{\type}{\operatorname{type}}

\hypersetup{
pdftitle={So Many Ways to Multiply 3x3 Matrices: A Large Dataset of Rank-23 Schemes},
pdfsubject={cs},
pdfauthor={Andrew Perminov},
pdfkeywords={Fast matrix multiplication, 3x3 matrix multiplication, rank 23, flip graph, dataset},
}

\begin{document}
\maketitle

\begin{abstract}
A publicly available dataset of \numSchemes{} rank-$23$ schemes for $3\times3$ matrix multiplication is presented. Every scheme in it has its own rank pattern, an invariant formed by the ranks of the factor matrices of a scheme, so all schemes are pairwise inequivalent. The schemes were obtained by a numerical method combining gradient descent with alternating least squares and by flip graph search, and have ternary, integer or rational coefficients. Rank $23$ is the best known for this format and is optimal for schemes with integer coefficients, but the number of essentially different schemes of this rank was poorly understood: the existing collections of Heule et al.\ and M{\aa}rtensson et al. contain many schemes that share only a small number of rank patterns (\heuleSchemes{} schemes with \heulePatterns{} patterns and \hellsBellsSchemes{} schemes with \hellsBellsPatterns{} patterns respectively). The dataset shows that the number of different rank patterns is much larger than previously known.
\end{abstract}


\input{structure/introduction}
\input{structure/preliminaries}
\input{structure/symmetries_and_invariants}
\input{structure/search_method}
\input{structure/dataset}
\input{structure/discussion}
\input{structure/conclusion}
\input{structure/availability}
\input{structure/acknowledgements}

\bibliographystyle{unsrtnat}
\bibliography{references}

\clearpage
\appendix
\input{structure/appendix}

\end{document}

%% file: structure/introduction.tex
\section{Introduction}
\label{sec:introduction}

Strassen's algorithm~\citep{strassen1969gaussian} multiplies two $2\times2$ matrices using seven multiplications instead of eight, and applied recursively it yields the exponent $\log_2 7 \approx 2.807$. Seven multiplications are optimal for this format~\citep{winograd1971multiplication, hopcroft1971minimizing}. The next square format, $3\times3$, is already considerably harder. Laderman's scheme~\citep{laderman1976noncommutative} uses $23$ multiplications, which corresponds to the exponent $\log_3 23 \approx 2.854$ and therefore does not beat Strassen's. Nevertheless, $3\times3$ remains the smallest square format for which the rank over fields of characteristic zero is unknown: it is only known to lie between $21$~\citep{Ivanov_A_lower_bound_2026} and $23$. The rank over $\mathbb{F}_2$ was recently shown to be exactly $23$~\citep{tomar2026rank}, which, by reduction modulo two, implies that $23$ is also optimal for schemes with integer coefficients.

Laderman's scheme is far from unique. Johnson and McLoughlin described parametrized families of rank-$23$ schemes~\citep{johnson1986noncommutative}, and Heule et al.~\citep{heule2021new} used SAT-based local search to obtain $\heuleSchemes$ pairwise inequivalent schemes with ternary coefficients. M{\aa}rtensson et al.~\citep{maartensson2026hell, fmm_add_reduction} later published a collection of \hellsBellsSchemes{} ternary schemes obtained with a neural network pipeline. In recent years, computer search has become a standard tool for small formats, based on reinforcement learning~\citep{fawzi2022discovering}, constraint programming~\citep{deza2023fast}, flip graphs~\citep{kauers2023flip, arai2024adaptive, moosbauer2025flip, kauers2025exploring, khoruzhii2026faster, perminov2025fast, perminov2026fast} and LLM-driven evolutionary search~\citep{novikov2025alphaevolve}. Catalogs of the best known schemes are maintained, for example, by Sedoglavic~\citep{sedoglavic2025yet}, Perminov~\citep{perminov2026fmm} and Lacelle~\citep{lacelle2026catalog}.

A large pool of rank-$23$ schemes is valuable because schemes of the same rank differ in properties that matter in practice and are not determined by the rank. The most prominent one is the number of additions required to evaluate a scheme. Laderman's original construction uses $98$ additions, and it can be reduced to $62$~\citep{martensson2024beast}. Further improvements, to $60$, $59$, $58$, $56$ and $55$ additions, as well as to $52$ and $51$ additions in an alternative basis, were obtained not for Laderman's scheme but for other rank-$23$ schemes~\citep{stapleton2025sixty, martensson2026fiftynine, perminov2025addition, sun2026fiftysix, karunaratne2026fiftyfive, sidebottom2026fiftyfive, nielsen202652, stapleton2026alternative}. The choice of the underlying scheme therefore matters, and a more diverse collection of schemes leaves more room for further improvements. Moreover, a diverse collection of decompositions provides opportunities for a more detailed investigation of the structure and properties of the matrix multiplication tensor.

This raises a natural question: \emph{how many essentially different rank-$23$ schemes exist?} Two schemes are considered equivalent if one can be transformed into the other by the symmetries of the matrix multiplication tensor (Section~\ref{sec:symmetries}). Heule et al.\ introduced invariants of this equivalence, the most basic of which is the \emph{rank pattern}: schemes with different rank patterns are necessarily inequivalent. Applied to the collection of~\citep{heule2021new, heule2019matrix}, these invariants reveal an interesting picture: its \heuleSchemes{} pairwise inequivalent schemes realize only \heulePatterns{} distinct rank patterns. The \hellsBellsSchemes{} schemes of~\citep{maartensson2026hell} realize only \hellsBellsPatterns{} patterns. In other words, in both collections many schemes share a rank pattern. This paper explores the opposite direction and shows that the number of different rank patterns is very large as well.

\paragraph{Contributions.}
\begin{itemize}
    \item A dataset of \numSchemes{} rank-$23$ schemes for $3\times3$ matrix multiplication is presented. The schemes were found by gradient descent with alternating least squares (GradALS), which produced more than half of them, and by flip graph search (over ternary coefficients and over $\mathbb{F}_2$ and $\mathbb{F}_3$ with Hensel lifting). Each scheme has its own rank pattern, hence all schemes are pairwise inequivalent. The coefficients are ternary for \countZT{} schemes (\percentZT\%), integer for \countZ{} (\percentZ\%) and rational for \countQ{} (\percentQ\%).

    \item The schemes of Heule et al.\ and M{\aa}rtensson et al.\ were not used as input to the search, yet all $\unionPatterns$ rank patterns of these two collections are present in the dataset. All other schemes of the dataset are inequivalent to every scheme of both collections. Together with the pairwise inequivalent schemes of Heule et al., the dataset gives at least \numOrbitsLB{} pairwise inequivalent rank-$23$ schemes (Section~\ref{sec:dataset-heule}).

    \item The distribution of the number of nonzero coefficients across the dataset is analyzed. A scheme with $138$ nonzero coefficients was found by GradALS method, whereas none of the previously known schemes from the publicly available collections had fewer than $139$. It uses the coefficients $\{0, \pm\nicefrac{1}{2}, \pm 1, \pm 2\}$. A scheme with the same number of nonzero coefficients and the same rank pattern was later published in a blog post~\citep{knottheory}. Since a ternary scheme with this rank pattern takes its place in the dataset, the scheme is given separately in Appendix~\ref{app:sparsest}.

    \item All schemes are available in JSON, plain text (\texttt{.txt}) and Maple (\texttt{.m}) formats together with a script that verifies their correctness. The repositories with the dataset and the tools are described in Section~\ref{sec:availability}.
\end{itemize}

The paper is organized as follows. Section~\ref{sec:preliminaries} fixes the notation, and Section~\ref{sec:symmetries} describes the symmetries and invariants used to distinguish schemes. Section~\ref{sec:search} explains how the schemes were found, Section~\ref{sec:dataset} describes the dataset, and Section~\ref{sec:discussion} discusses the results and limitations. Sections~\ref{sec:conclusion} and~\ref{sec:availability} contain the conclusion and the description of the available data and code, and Appendix~\ref{app:sparsest} contains the sparsest scheme.

%% file: structure/preliminaries.tex
\section{Preliminaries}
\label{sec:preliminaries}

\subsection{Matrix multiplication schemes}

Let $\mathbb{F}$ be a field and let $A, B \in \mathbb{F}^{n \times n}$ be two matrices with entries $a_{i,j}$ and $b_{i,j}$. A \emph{scheme} of rank $r$ for the multiplication of $n \times n$ matrices, denoted $(n,n,n\!:\!r)$, is a way to compute the entries of $C = AB$ using $r$ multiplications of linear forms:
\begin{align*}
    m_l &= \Bigl(\sum_{i,j=1}^{n} u^{(l)}_{i,j}\, a_{i,j}\Bigr) \cdot \Bigl(\sum_{i,j=1}^{n} v^{(l)}_{i,j}\, b_{i,j}\Bigr), \qquad l = 1, \dots, r, \\
    c_{i,j} &= \sum_{l=1}^{r} w^{(l)}_{j,i}\, m_l, \qquad i, j = 1, \dots, n,
\end{align*}
where $u^{(l)}_{i,j}, v^{(l)}_{i,j}, w^{(l)}_{i,j} \in \mathbb{F}$.

\begin{notation}
The coefficients are collected in three matrices $U, V, W \in \mathbb{F}^{r \times n^2}$, whose entries are indexed in the row-major order of the pairs $(i,j)$:
\[
    u^{(l)}_{i,j} = U_{l,\,(i-1)n + j}, \qquad v^{(l)}_{i,j} = V_{l,\,(i-1)n + j}, \qquad w^{(l)}_{i,j} = W_{l,\,(i-1)n + j}.
\]
The symbols $u^{(l)}$, $v^{(l)}$ and $w^{(l)}$ without subscripts denote the $l$-th rows of $U$, $V$ and $W$, which are also regarded as $n \times n$ matrices with entries $u^{(l)}_{i,j}$, $v^{(l)}_{i,j}$ and $w^{(l)}_{i,j}$. The rank of such a matrix is denoted by $\rk{u^{(l)}}$ to distinguish it from the rank $r$ of the scheme.
\end{notation}

Let $\operatorname{vec}$ denote the vectorization of a matrix in the row-major order. Then the scheme computes
\begin{equation}
    \operatorname{vec}(C^{T}) = W^{T} \bigl( (U \operatorname{vec}(A)) \odot (V \operatorname{vec}(B)) \bigr),
    \label{eq:scheme-matrix-form}
\end{equation}
where $\odot$ is the elementwise (Hadamard) product. In other words, the scheme is fully described by the triple $(U, V, W)$.

\begin{remark*}
The matrix $W$ is stored in the transposed form: $w^{(l)}_{i,j}$ is the coefficient with which $m_l$ enters the entry $c_{j,i}$ of the product rather than $c_{i,j}$. This corresponds to the representation of matrix multiplication via the trace form $\tr{(ABC)}$, and it is the encoding used by most catalogs of schemes as well as by the files of the dataset described in Section~\ref{sec:dataset}. As shown below, it also makes all three matrices play symmetric roles.
\end{remark*}

\begin{example}
Strassen's original scheme~\citep{strassen1969gaussian} is a scheme $(2,2,2\!:\!7)$ over $\mathbb{Z}$:
\begin{align*}
    m_1 &= (a_{1,1} + a_{2,2})(b_{1,1} + b_{2,2}) & c_{1,1} &= m_1 + m_4 - m_5 + m_7 \\
    m_2 &= (a_{2,1} + a_{2,2})\, b_{1,1}          & c_{1,2} &= m_3 + m_5 \\
    m_3 &= a_{1,1} (b_{1,2} - b_{2,2})            & c_{2,1} &= m_2 + m_4 \\
    m_4 &= a_{2,2} (b_{2,1} - b_{1,1})            & c_{2,2} &= m_1 - m_2 + m_3 + m_6 \\
    m_5 &= (a_{1,1} + a_{1,2})\, b_{2,2} \\
    m_6 &= (a_{2,1} - a_{1,1})(b_{1,1} + b_{1,2}) \\
    m_7 &= (a_{1,2} - a_{2,2})(b_{2,1} + b_{2,2})
\end{align*}

\end{example}

\subsection{Brent equations}

A triple $(U, V, W)$ defines a scheme if and only if its coefficients satisfy the Brent equations~\citep{brent1970algorithms}:
\begin{equation}
    \sum_{l=1}^{r} u^{(l)}_{i_1,i_2}\, v^{(l)}_{j_1,j_2}\, w^{(l)}_{k_1,k_2} = \delta_{i_2 j_1}\, \delta_{j_2 k_1}\, \delta_{k_2 i_1}
    \label{eq:brent}
\end{equation}
for all indices $1 \le i_1, i_2, j_1, j_2, k_1, k_2 \le n$, where $\delta_{ij}$ is the Kronecker delta, equal to $1$ if $i = j$ and to $0$ otherwise.

\subsection{Tensor representation}

\begin{definition}[\textbf{Matrix multiplication tensor}]
The matrix multiplication tensor $T_{n,n,n} \in \mathbb{F}^{n^2 \times n^2 \times n^2}$ is the three-dimensional tensor with entries
\[
    (T_{n,n,n})_{(i_1 - 1)n + i_2,\; (j_1 - 1)n + j_2,\; (k_1 - 1)n + k_2} = \delta_{i_2 j_1}\, \delta_{j_2 k_1}\, \delta_{k_2 i_1}.
\]
\end{definition}

\begin{definition}[\textbf{CP decomposition}]
A decomposition of a tensor $T$ into a sum of $r$ rank-one tensors is an expression
\[
    T = \sum_{l=1}^{r} u^{(l)} \otimes v^{(l)} \otimes w^{(l)},
\]
where $\otimes$ denotes the outer product. The smallest $r$ for which such a decomposition exists is the \emph{tensor rank} of $T$.
\end{definition}

Comparing the right-hand side of~\eqref{eq:brent} with the definition of $T_{n,n,n}$ shows that schemes of rank $r$ for $n \times n$ matrices are exactly the decompositions of $T_{n,n,n}$ into $r$ rank-one tensors, and that the rank of the scheme is the length of the decomposition.

\begin{example}
Strassen's scheme from the previous example corresponds to the decomposition
\begin{align*}
    T_{2,2,2} ={}& (a_{1,1} + a_{2,2}) \otimes (b_{1,1} + b_{2,2}) \otimes (c_{1,1} + c_{2,2}) \\
    +\;& (a_{2,1} + a_{2,2}) \otimes (b_{1,1}) \otimes (c_{1,2} - c_{2,2}) \\
    +\;& (a_{1,1}) \otimes (b_{1,2} - b_{2,2}) \otimes (c_{2,1} + c_{2,2}) \\
    +\;& (a_{2,2}) \otimes (b_{2,1} - b_{1,1}) \otimes (c_{1,1} + c_{1,2}) \\
    +\;& (a_{1,1} + a_{1,2}) \otimes (b_{2,2}) \otimes (c_{2,1} - c_{1,1}) \\
    +\;& (a_{2,1} - a_{1,1}) \otimes (b_{1,1} + b_{1,2}) \otimes (c_{2,2}) \\
    +\;& (a_{1,2} - a_{2,2}) \otimes (b_{2,1} + b_{2,2}) \otimes (c_{1,1}).
\end{align*}

Because of the transposed encoding of $W$, the third factor of each term is written in the basis $c_{j,i}$: for instance, in Strassen's scheme $m_2$ enters $c_{2,1}$ and $-c_{2,2}$, and the third factor of the second term is $c_{1,2} - c_{2,2}$.
\end{example}

\subsection{Coefficient classes and sparsity}

The coefficients of the schemes considered in this paper are rational numbers. Depending on the set of values they take, the schemes are divided into three classes: \emph{ternary} schemes with coefficients in $\mathbb{Z}_T = \{-1, 0, 1\}$, \emph{integer} schemes with coefficients in $\mathbb{Z}$, and \emph{rational} schemes with coefficients in $\mathbb{Q}$. The class of a scheme is the smallest one that contains all its coefficients, so the classes are ordered as $\mathbb{Z}_T \subset \mathbb{Z} \subset \mathbb{Q}$. Ternary schemes are the most convenient: they work over any field and require no multiplications by constants.

The \emph{sparsity} of a scheme is measured by the total number of its nonzero coefficients $\operatorname{nnz}(U) + \operatorname{nnz}(V) + \operatorname{nnz}(W)$. The fewer nonzero coefficients a scheme has, the simpler are the linear forms it computes, and the fewer (naive) additions are needed to evaluate it directly.

%% file: structure/symmetries_and_invariants.tex
\section{Symmetries and invariants}
\label{sec:symmetries}

Matrix multiplication has a number of symmetries, which allow one to obtain new schemes from a known one. Each of the following transformations maps a scheme of rank $r$ to a scheme of the same rank:
\begin{enumerate}
    \item[(a)] \emph{Reordering of terms}: the triples $(u^{(l)}, v^{(l)}, w^{(l)})$ are permuted arbitrarily.
    \item[(b)] \emph{Rescaling}: $(u^{(l)}, v^{(l)}, w^{(l)}) \mapsto (\alpha u^{(l)}, \beta v^{(l)}, \gamma w^{(l)})$ for nonzero $\alpha, \beta, \gamma \in \mathbb{F}$ with $\alpha\beta\gamma = 1$. The constants may be chosen separately for every term, since the product of the three factors does not change.
    \item[(c)] \emph{Sandwiching}: $(u^{(l)}, v^{(l)}, w^{(l)}) \mapsto (P u^{(l)} Q^{-1},\; Q v^{(l)} R^{-1},\; R w^{(l)} P^{-1})$ for all $l$, where $P, Q, R \in \mathbb{F}^{n \times n}$ are invertible matrices. It is based on the identity $(P A Q^{-1})(Q B R^{-1}) = P (AB) R^{-1}$: a scheme can be applied to the transformed matrices, and the result can then be transformed back.
    \item[(d)] \emph{Cyclic shift}: $(U, V, W) \mapsto (V, W, U)$.
    \item[(e)] \emph{Transposition}: $(u^{(l)}, v^{(l)}, w^{(l)}) \mapsto ((v^{(l)})^{T}, (u^{(l)})^{T}, (w^{(l)})^{T})$ for all $l$. It reflects the identity $(AB)^{T} = B^{T} A^{T}$.
\end{enumerate}
Transformations (d) and (e) are a consequence of the transposed encoding of $W$, which makes the three matrices $U$, $V$ and $W$ play symmetric roles: the matrix multiplication tensor corresponds to the trilinear form $\tr(ABC)$, which does not change under the cyclic shift, $\tr(ABC) = \tr(CAB)$, and under the transposition, $\tr(ABC) = \tr(C^{T} B^{T} A^{T})$.

Transformations (c)--(e) are symmetries of the tensor $T_{n,n,n}$ itself: sandwiching is the action of the group $\mathrm{GL}_n^{\times 3}$ studied by de~Groote~\citep{degroote1978varieties}, while (d) and (e) permute the three tensor factors (in the case of (e), together with transposing the matrices $u^{(l)}, v^{(l)}, w^{(l)}$) and generate a group isomorphic to the symmetric group $S_3$~\citep{hopcroft1973duality}. Together they generate the full symmetry group of $T_{n,n,n}$~\citep{degroote1978varieties, burichenko2014symmetries}. Transformations (a) and (b), in contrast, are trivial symmetries of the representation of a tensor as a sum of rank-one terms.

\begin{definition}[\textbf{Equivalence of schemes}]
Two schemes are \emph{equivalent} if one can be obtained from the other by a finite sequence of transformations (a)--(e). The set of schemes equivalent to a given one is called its \emph{orbit}.
\end{definition}

Neither the coefficient class nor the number of nonzero coefficients is preserved by these transformations: sandwiching with a suitable matrix can turn a ternary scheme into an integer or rational one, and vice versa. Sandwiching is therefore a way to explore the orbit of a scheme and to look for sparser or ``nicer'' representatives in it, which is how it is used in the search described in Section~\ref{sec:search}.

For $n = 2$, all schemes of rank $7$ form a single orbit, which is the orbit of Strassen's scheme~\citep{degroote1978varieties}. For $n = 3$, the situation is completely different: there is a great number of orbits of rank-$23$ schemes, and telling them apart is the topic of the rest of this section.

\subsection{Rank pattern}
\label{sec:rank-pattern}

A convenient way to distinguish orbits is to look at the ranks of the matrices $u^{(l)}$, $v^{(l)}$ and $w^{(l)}$, as proposed by Heule et al.~\citep{heule2021new}. For a scheme with the terms $(u^{(l)}, v^{(l)}, w^{(l)})$, consider the triples
\[
    \rho_l = \bigl(\rk u^{(l)},\, \rk v^{(l)},\, \rk w^{(l)}\bigr), \qquad l = 1, \dots, r,
\]
and apply the same permutation of coordinates to all of them. There are six such permutations (including the identity), and each of them gives a list of $r$ triples, which is then sorted in the lexicographic order.

\begin{definition}[\textbf{Rank pattern}]
The \emph{rank pattern} of a scheme is the lexicographically smallest of the six sorted lists described above.
\end{definition}

\begin{example}
Laderman's scheme has the following sequence of ranks $\rho_l$:
\[
\begin{array}{cccccccccccccccccccccccc}
\rk u^{(l)} & 3 & 1 & 1 & 2 & 1 & 1 & 2 & 1 & 1 & 3 & 1 & 2 & 1 & 1 & 1 & 2 & 1 & 1 & 1 & 1 & 1 & 1 & 1 \\
\rk v^{(l)} & 1 & 1 & 3 & 2 & 1 & 1 & 2 & 1 & 1 & 1 & 3 & 2 & 1 & 1 & 1 & 2 & 1 & 1 & 1 & 1 & 1 & 1 & 1 \\
\rk w^{(l)} & 1 & 1 & 1 & 2 & 1 & 3 & 2 & 1 & 1 & 1 & 1 & 2 & 1 & 3 & 1 & 2 & 1 & 1 & 1 & 1 & 1 & 1 & 1 \\
\end{array}
\]
After sorting in lexicographic order and choosing the smallest of the six permutations, this gives the rank pattern
\[
\begin{array}{ccccccccccccccccccccccc}
1 & 1 & 1 & 1 & 1 & 1 & 1 & 1 & 1 & 1 & 1 & 1 & 1 & 1 & 1 & 1 & 1 & 2 & 2 & 2 & 2 & 3 & 3 \\
1 & 1 & 1 & 1 & 1 & 1 & 1 & 1 & 1 & 1 & 1 & 1 & 1 & 1 & 1 & 3 & 3 & 2 & 2 & 2 & 2 & 1 & 1 \\
1 & 1 & 1 & 1 & 1 & 1 & 1 & 1 & 1 & 1 & 1 & 1 & 1 & 3 & 3 & 1 & 1 & 2 & 2 & 2 & 2 & 1 & 1 \\
\end{array}
\]
\end{example}

Equivalent schemes have the same rank pattern. Indeed, matrix rank does not change under reordering of terms, rescaling and sandwiching, so these transformations preserve the multiset of triples $\rho_l$. The cyclic shift and the transposition permute the coordinates of every triple in the same way (a matrix and its transpose have equal ranks), and the two of them generate all six permutations, which are taken into account by the choice of the smallest list. Consequently, schemes with different rank patterns are necessarily inequivalent.

The converse is not true: inequivalent schemes may share a rank pattern. In particular, the \heuleSchemes{} pairwise inequivalent schemes of~\citep{heule2021new} realize only \heulePatterns{} different rank patterns. Equality of rank patterns is thus a necessary condition for equivalence, but not a sufficient one. An algorithm that decides whether two schemes are equivalent was proposed in~\citep{kauers2022normal}, but it is not used in this work: the statement about different rank patterns is the only tool used here to claim that two schemes are inequivalent. The dataset described in Section~\ref{sec:dataset} contains exactly one scheme for every rank pattern, which makes all its schemes pairwise inequivalent. In other words, the dataset contains one representative per rank pattern, and not one representative per orbit.

\subsection{Polynomial invariants and the type of a scheme}
\label{sec:polynomial-invariants}

The rank pattern can be conveniently encoded by a polynomial. For a scheme with the triples $\rho_l$ defined above, let
\begin{equation}
    \type(x, y, z) = \sum_{l=1}^{r} x^{\rk u^{(l)}} y^{\rk v^{(l)}} z^{\rk w^{(l)}},
\end{equation}
which is the invariant used in Sedoglavic's catalog~\citep{sedoglavic2025yet}. Reordering, rescaling and sandwiching do not change it, while the cyclic shift and the transposition permute the variables $x, y, z$.

Heule et al.~\citep{heule2021new} additionally introduced two polynomial invariants, which carry less information than the rank pattern but are much easier to tabulate and compare:
\begin{align}
    f(x, y, z) &= \sum_{\pi \in S_3} \pi\left(\sum_{l=1}^{r} x^{\rk u^{(l)}} y^{\rk v^{(l)}} z^{\rk w^{(l)}}\right), \\
    g(\omega) &= \omega^{\sum_{l=1}^{r} \rk u^{(l)}} + \omega^{\sum_{l=1}^{r} \rk v^{(l)}} + \omega^{\sum_{l=1}^{r} \rk w^{(l)}},
\end{align}
where the permutations $\pi$ are meant to permute the variables $x$, $y$ and $z$. Both polynomials do not change under any of the transformations (a)--(e).

\begin{example}
For Laderman's scheme, the type, the polynomials $f$ and $g$ take the following forms:
\[
\type(x,y,z) = 4x^2y^2z^2 + 2x^3yz + 2xy^3z + 2xyz^3 + 13xyz.
\]
\[f(x,y,z) = 24x^2y^2z^2 + 12x^3yz + 12xy^3z + 12xyz^3 + 78xyz,
\]
and
\[
g(\omega) = 3\omega^{31}.
\]
\end{example}

The polynomial $\type$ itself depends on the order of the three factors, so every scheme is brought to a \emph{sorted form}: the matrices $U$, $V$, $W$ are permuted by (d) and (e) so that the sorted list of triples $\rho_l$ is exactly the rank pattern, and the terms are sorted in the increasing lexicographic order of these triples. The order of terms with equal triples, and the choice among permutations giving the same list, is not fixed, so the sorted form is not unique. Unlike the normal form of~\citep{kauers2022normal}, it does not single out a representative of an orbit and only makes the rank pattern readable directly from the files. All schemes of the dataset are stored in this form. The polynomial $\type$ of a scheme in sorted form encodes its rank pattern one-to-one, so the number of distinct types in the dataset equals the number of distinct rank patterns, while the numbers of distinct values of $f$ and $g$ are smaller, since these invariants are coarser. All three quantities are used in Section~\ref{sec:dataset} to compare the dataset with the collections of Heule et al.\ and M{\aa}rtensson et al.

%% file: structure/search_method.tex
\section{Search method}
\label{sec:search}

The schemes of the dataset were obtained by two groups of methods: the numerical method GradALS (Section~\ref{sec:als}), which produced more than half of the dataset, including the vast majority of rational schemes, and the flip graph search, performed over ternary coefficients and over finite fields with Hensel lifting (Sections~\ref{sec:flip-graphs}--\ref{sec:coefficient-sets}). The schemes found by either group are compared with the same pool of known schemes (Section~\ref{sec:search-procedure}).

\subsection{Flip graphs}
\label{sec:flip-graphs}

A part of the schemes was obtained with the flip graph approach~\citep{kauers2023flip}. In a flip graph, every vertex is a scheme (a decomposition of the tensor $T_{n,n,n}$ into rank-one terms), and edges correspond to transformations that preserve correctness. Since the graph is far too large to be explored exhaustively, it is searched by random walks. Three kinds of transformations are used, each of them being applied to a pair of terms $u^{(i)} \otimes v^{(i)} \otimes w^{(i)}$ and $u^{(j)} \otimes v^{(j)} \otimes w^{(j)}$.

\begin{definition}[\textbf{Flip}]
If $u^{(i)} = u^{(j)}$, then
\begin{align*}
    u^{(i)} \otimes v^{(i)} \otimes w^{(i)} \;+\; &u^{(j)} \otimes v^{(j)} \otimes w^{(j)} \;\rightarrow \\
    u^{(i)} \otimes (v^{(i)} + v^{(j)}) \otimes w^{(i)} \;+\; &u^{(j)} \otimes v^{(j)} \otimes (w^{(j)} - w^{(i)}).
\end{align*}
The same transformation is defined for $v^{(i)} = v^{(j)}$ and for $w^{(i)} = w^{(j)}$ by permuting the roles of the three factors. A flip preserves both the correctness and the rank of a scheme.
\end{definition}

\begin{definition}[\textbf{Reduction}]
If $u^{(i)} = u^{(j)}$ and $v^{(i)} = v^{(j)}$, then
\[
    u^{(i)} \otimes v^{(i)} \otimes w^{(i)} \;+\; u^{(j)} \otimes v^{(j)} \otimes w^{(j)} \;\rightarrow\; u^{(i)} \otimes v^{(i)} \otimes (w^{(i)} + w^{(j)}).
\]
The same holds for equalities in any other pair of factors. A reduction decreases the rank by one. The rank also decreases when a flip produces a term with a zero factor, and such a term is simply removed.
\end{definition}

\begin{definition}[\textbf{Plus}]
If $u^{(i)} \neq u^{(j)}$, $v^{(i)} \neq v^{(j)}$ and $w^{(i)} \neq w^{(j)}$, then
\begin{align*}
    u^{(i)} \otimes v^{(i)} \otimes w^{(i)} \;+\; &u^{(j)} \otimes v^{(j)} \otimes w^{(j)} \;\rightarrow \\
    u^{(i)} \otimes (v^{(i)} + v^{(j)}) \otimes w^{(i)} \;+\; &u^{(j)} \otimes v^{(j)} \otimes (w^{(j)} - w^{(i)}) \;+\; (u^{(j)} - u^{(i)}) \otimes v^{(j)} \otimes w^{(i)}.
\end{align*}
This transformation increases the rank by one and preserves correctness.
\end{definition}

Flips alone never change the rank, and a scheme without equal factors in different terms has no flips available. The plus transformation~\citep{arai2024adaptive} helps to leave such dead ends: it increases the rank by one and creates new coincidences between the factors, after which reductions can bring the rank back. Since flips and reductions both produce schemes that remain correct, every visited scheme of rank $23$ is a valid scheme for $3 \times 3$ matrices.

\subsection{Search procedure}
\label{sec:search-procedure}

The pool of schemes coincides with the dataset: it stores exactly one scheme for every rank pattern found so far by any of the methods. The flip graph search is performed by a number of independent runners, each of which repeats the following steps (Algorithm~\ref{alg:search}) starting from the initial pool of rank-$23$ schemes obtained in earlier flip graph searches~\citep{perminov2025fast, perminov2026fast}:
\begin{enumerate}
    \item A scheme is chosen from the pool at random.
    \item A random walk in the flip graph is started from it. Most of the steps are flips, and reductions are applied whenever they are available. From time to time, a plus transformation is applied with some probability, which allows the walk to leave the neighbourhood of the current scheme.
    \item Periodically, a random sandwiching (Section~\ref{sec:symmetries}) is applied to the current scheme. It does not change the rank pattern, but it changes the coefficients and often helps to obtain sparser schemes, that is, schemes with fewer nonzero coefficients.
    \item Each visited scheme of rank $23$ is compared with the pool as described below.
\end{enumerate}

\begin{algorithm}[t]
\caption{Flip graph search used to build the dataset}
\label{alg:search}
\KwIn{pool $\mathcal{D}$ of known rank-$23$ schemes, one per rank pattern}
\KwOut{updated pool $\mathcal{D}$}
\While{the search is not stopped}{
    $S \gets$ random scheme from $\mathcal{D}$\;
    \Repeat{walk length is reached}{
        apply a flip to $S$, or a plus with a small probability\;
        apply all available reductions to $S$\;
        \If{the sandwiching step is due}{
            apply a random sandwiching to $S$\;
        }
        \If{$\operatorname{rank}(S) = 23$}{
            $p \gets$ rank pattern of $S$\;
            \lIf{$p$ is not in $\mathcal{D}$}{add $S$ to $\mathcal{D}$}
            \ElseIf{$S$ is better than the scheme with pattern $p$ in $\mathcal{D}$}{
                replace that scheme by $S$\;
            }
        }
    }
}
\end{algorithm}

A scheme with a new rank pattern is added to the pool. If a scheme with an already known pattern is encountered, it replaces the stored one in either of two cases: its coefficients come from a better class, in the order $\mathbb{Z}_T \succ \mathbb{Z} \succ \mathbb{Q}$ (Section~\ref{sec:preliminaries}), or the class is the same and the scheme has fewer nonzero coefficients. The class takes priority over sparsity, so a ternary scheme always replaces an integer or rational one, even if the latter is sparser. As a result, the final dataset contains, for every rank pattern, the scheme with the best coefficient class and the smallest number of nonzero coefficients among those found during the search.

This rule has two consequences that should be kept in mind. First, the minimal possible number of nonzero coefficients was not the goal: the rule is used to avoid keeping the first scheme found for a pattern, which may be unnecessarily dense, but sparser schemes with the same rank pattern may well exist. Second, integer and rational schemes of the dataset are those for which no ternary scheme with the same pattern has been found yet, and such a scheme may still exist.

\subsection{Coefficient sets: ternary search and Hensel lifting}
\label{sec:coefficient-sets}

The flip graph search was performed mainly in the ternary flip graph~\citep{perminov2025fast}, in which all coefficients of the schemes belong to $\mathbb{Z}_T = \{-1, 0, 1\}$. Transformations producing coefficients outside this set are not performed, so every visited scheme is ternary.

In addition, the flip graph search was performed over the fields $\mathbb{F}_2$ and $\mathbb{F}_3$. Over $\mathbb{F}_2$, all coefficients are $0$ or $1$ and signs disappear, whereas over $\mathbb{F}_3$ the values $1$ and $-1$ remain distinct. A scheme found over $\mathbb{F}_p$, where $p \in \{2, 3\}$, is lifted to characteristic zero by Hensel lifting: starting from a solution of the Brent equations modulo $p$, the solution is refined step by step to a solution modulo a higher power of $p$. At most three lifting steps are performed for $p = 2$ and a single step for $p = 3$. After the last step, every coefficient is reconstructed from its residue as a rational number (rational reconstruction), and the resulting triple $(U, V, W)$ is accepted if it satisfies the Brent equations exactly. Schemes that do not admit such a reconstruction within the allowed number of steps are discarded, so only lifts with small numerators and denominators are kept.

The class of a lifted scheme depends on the reconstructed values. If all coefficients lie in $\mathbb{Z}_T$, the scheme is ternary. If they are integers and some of them lie outside $\mathbb{Z}_T$ (for example, $2$), the scheme is integer. Otherwise, the reconstruction produces fractions, whose denominators are not divisible by $p$, and the scheme is rational.

\subsection{Gradient descent with alternating least squares (GradALS)}
\label{sec:als}

More than half of the schemes of the dataset were obtained by a numerical method, referred to as GradALS, which combines gradient descent with alternating least squares (ALS). The method approximates the tensor $T_{3,3,3}$ by a sum of $23$ rank-one terms. The approximation error is minimized by gradient descent, and ALS steps, which alternately solve linear least squares problems for the matrices $U$, $V$ and $W$ while the other two are fixed, are used to speed up convergence. Several regularizers push the coefficients towards the set $\{0, \pm\nicefrac{1}{2}, \pm1, \pm2\}$. The numerical solutions obtained are converted to exact schemes and verified. Any subset of this set of values may occur in a solution, so GradALS produces ternary schemes, integer schemes (with the coefficient $\pm2$) and rational schemes with denominator $2$, and the class of each scheme is determined by the values actually obtained. The vast majority of rational schemes of the dataset were obtained in this way, rather than by the flip graph search over $\mathbb{F}_3$ with lifting. A detailed description of the method will be published separately, and its implementation is available in the repository described in Section~\ref{sec:availability}.

%% file: structure/dataset.tex
\section{The dataset}
\label{sec:dataset}

\subsection{Contents}
\label{sec:dataset-contents}

The dataset consists of \numSchemes{} schemes of rank $23$ for the multiplication of $3 \times 3$ matrices. Every scheme is given by the matrices $U$, $V$, $W \in \mathbb{F}^{23 \times 9}$ in the encoding of Section~\ref{sec:preliminaries}. All schemes are stored in the sorted form of Section~\ref{sec:polynomial-invariants}: the matrices are permuted to match the rank pattern, and their rows are sorted in the increasing lexicographic order of the triples of ranks.

For every rank pattern, the dataset contains exactly one scheme, which is selected by the rule of Section~\ref{sec:search-procedure}: it has the best coefficient class and, within this class, the smallest number of nonzero coefficients among the schemes of this pattern found during the search. As a consequence, all schemes of the dataset are pairwise inequivalent. At the same time, the dataset is not complete: it neither contains all orbits of rank-$23$ schemes nor, in all likelihood, all possible rank patterns, and several inequivalent schemes may exist for each of its patterns.

By the coefficient classes of Section~\ref{sec:preliminaries}, \countZT{} schemes (\percentZT\%) are ternary, \countZ{} (\percentZ\%) are integer, and \countQ{} (\percentQ\%) are rational. Integer and rational schemes are those for which no ternary scheme with the same rank pattern has been found.

\subsection{Statistics of the invariants}
\label{sec:dataset-invariants}

Table~\ref{tab:invariants} shows how many distinct values of the invariants of Section~\ref{sec:symmetries} are realized by the schemes of the dataset. Since every scheme has its own rank pattern, the number of distinct types coincides with the number of schemes. The polynomials $f$ and $g$ are coarser invariants, so they take considerably fewer values. The numbers in the columns of individual classes are counted separately within each class, and they do not add up to the total except for the row of types, since the same value of $f$ or $g$ may occur in schemes of different classes.

\begin{table}[th]
\centering
\caption{Number of distinct values of the invariants in the dataset, in total and separately for each coefficient class.}
\label{tab:invariants}
\begin{tabular}{lrrrr}
\toprule
Invariant & All & Ternary & Integer & Rational \\
\midrule
$\type(x, y, z)$ & \numSchemes & \countZT & \countZ & \countQ \\
$f(x, y, z)$ & \numF & \numFZT & \numFZ & \numFQ \\
$g(\omega)$ & \numG & \numGZT & \numGZ & \numGQ \\
\bottomrule
\end{tabular}
\end{table}

The coarse invariants show a different picture from the counts of schemes. The three coefficient classes are of comparable size (\percentZT\%, \percentZ\% and \percentQ\% of the dataset), yet ternary schemes realize the fewest distinct values of $f$ and $g$: \numFZT{} and \numGZT, against \numFZ{} and \numGZ{} for integer schemes and \numFQ{} and \numGQ{} for rational schemes. This indicates that the non-ternary part of the dataset is not concentrated in a few rank patterns, but is spread over a wide range of them.

\subsection{Sparsity}
\label{sec:dataset-sparsity}

\begin{figure}[th]
\centering
\includegraphics[width=\linewidth]{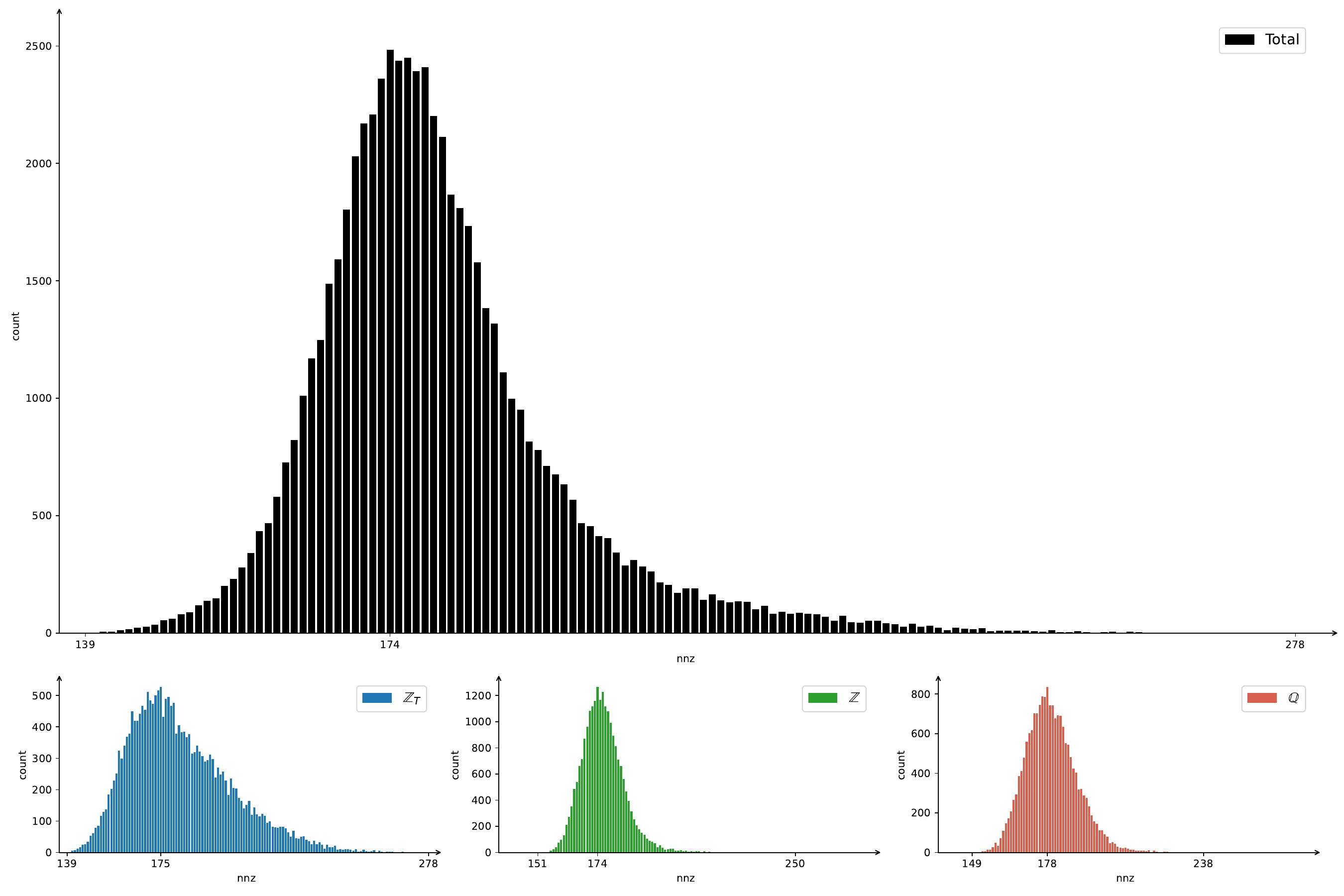}%
\caption{Distribution of the number of nonzero coefficients over all schemes of the dataset (top) and separately over the ternary ($\mathbb{Z}_T$), integer ($\mathbb{Z}$) and rational ($\mathbb{Q}$) schemes (bottom). The ticks on the horizontal axes mark the minimum, the mode and the maximum of each distribution; the vertical scales of the panels differ.}
\label{fig:nnz}
\end{figure}

The sparsity of a scheme is measured by the total number of nonzero coefficients in $U$, $V$ and $W$. Figure~\ref{fig:nnz} shows the distribution of this number over the dataset.

The distribution is unimodal and right-skewed: most schemes lie in a narrow window around the peak, the sparse end, starting from $\minNnz$, is thinly populated, and the dense side extends into a long tail up to $\maxNnz$. Schemes close to the minimum are therefore rare.

The coefficient classes differ mainly in spread rather than in location. The ternary schemes span the whole range, and they are the only ones in its sparsest part: every scheme with fewer than $\minNnzNonZT$ nonzero coefficients is ternary. The integer and rational schemes are concentrated in narrower windows around nearly the same peak: the modes are $\modeNnzZT$, $\modeNnzZ$ and $\modeNnzQ$ for ternary, integer and rational schemes, respectively.

These distributions describe the dataset, and not all rank-$23$ schemes. By the selection rule of Section~\ref{sec:search-procedure}, a scheme of a more general class is kept only for a rank pattern without a ternary scheme, and a ternary scheme is preferred regardless of the number of nonzero coefficients, so a sparser scheme with the same rank pattern may exist outside the dataset. The scheme with $138$ nonzero coefficients described in Section~\ref{sec:dataset-sparsest} is an example: it is rational, and the ternary scheme that represents its pattern in the dataset has more nonzero coefficients.

\subsection{Rank triples}
\label{sec:dataset-triples}

\begin{figure}[t]
    \centering
    \includegraphics[width=\textwidth]{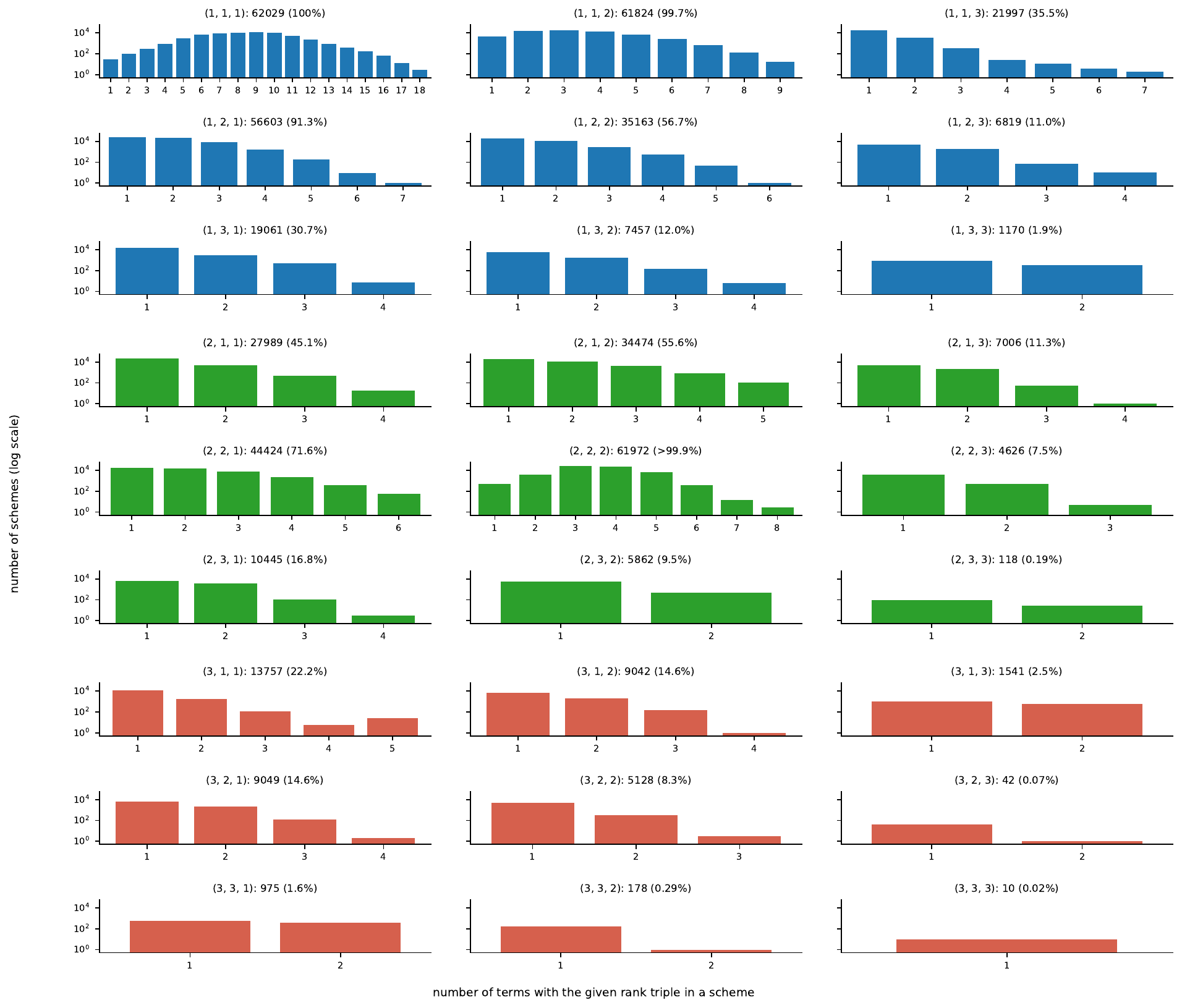}
    \caption{Distribution of the number of terms with a given rank triple $(\rk u^{(l)}, \rk v^{(l)}, \rk w^{(l)})$ in a scheme, for each of the $27$ triples. The horizontal axis shows the number of such terms in a scheme, and the vertical axis (logarithmic) shows the number of schemes of the dataset. The percentage in the title of a panel is the share of schemes that contain the triple at least once; schemes without it are not shown. The colors correspond to the rank of the first factor.}
    \label{fig:triples}
\end{figure}

The rank pattern of a scheme is the multiset of the triples $\rho_l = (\rk u^{(l)}, \rk v^{(l)}, \rk w^{(l)})$ of its $23$ terms, so a scheme can be seen as a way of distributing its terms among the $27$ possible triples. Figure~\ref{fig:triples} shows how the schemes of the dataset do this: for every triple, it gives the distribution of the number of terms of that type.

Several features of this picture explain why rank patterns are so numerous. Terms with factors of low rank dominate, and the triples $(1,1,1)$, $(1,1,2)$ and $(2,2,2)$ are present in practically every scheme, so they form a common core of rank-$23$ schemes. The variety lies in how many terms of each type a scheme has: for $(1,1,1)$ this number varies over a very wide range, and for most other frequent triples it takes several values. A rank pattern is thus a combination of several counts that vary almost independently, and this combinatorial freedom accounts for most of the patterns. At the other end, triples with at least two factors of rank $3$ are exceptional: they occur in a small fraction of schemes, and at most twice in a scheme.

The panels are not symmetric with respect to permutations of a triple, for example $(1,1,2)$ is far more common than $(2,1,1)$. This is a consequence of the sorted form (Section~\ref{sec:polynomial-invariants}): the order of the factors of every scheme is chosen so that its list of triples is lexicographically smallest, which favors triples with small leading entries. The figure therefore describes the rank patterns as they are defined, and not the ranks of individual factors.

\subsection{Comparison with existing collections}
\label{sec:dataset-heule}

Table~\ref{tab:collections} compares the dataset with two existing collections of rank-$23$ schemes: that of Heule et al.~\citep{heule2021new}, which is the closest to it in nature, and that of M{\aa}rtensson et al.~\citep{maartensson2026hell}. The first collection contains \heuleSchemes{} pairwise inequivalent schemes with ternary coefficients, which realize only \heulePatterns{} distinct rank patterns and \heuleF{} distinct values of $f$. The second one contains \hellsBellsSchemes{} schemes with ternary coefficients, which realize \hellsBellsPatterns{} rank patterns and \hellsBellsF{} values of $f$. Their pairwise inequivalence was not established. The two collections have \commonPatterns{} rank patterns in common. The dataset, in contrast, contains \numSchemes{} schemes with pairwise distinct rank patterns and \numF{} distinct values of $f$.

\begin{table}[t]
\centering
\caption{Comparison with existing collections of rank-$23$ $3\times3$ schemes. The last two columns show the ternary part of the dataset and the whole dataset.}
\label{tab:collections}
\begin{tabular}{lrrrr}
\toprule
 & Heule et al. & M{\aa}rtensson et al. & \shortstack{This dataset\\(ternary)} & \shortstack{This dataset\\(full)} \\
\midrule
Schemes & \heuleSchemes & \hellsBellsSchemes & \countZT & \numSchemes \\
Pairwise inequivalent & yes & unknown & yes & yes \\
Distinct rank patterns & \heulePatterns & \hellsBellsPatterns & \countZT & \numSchemes \\
Distinct values of $f(x,y,z)$ & \heuleF & \hellsBellsF & \numFZT & \numF \\
Distinct values of $g(\omega)$ & \heuleG & \hellsBellsG & \numGZT & \numG \\
Nonzero coefficients (range) & \heuleMinNnz-\heuleMaxNnz & \hellsBellsMinNnz-\hellsBellsMaxNnz & \minNnz-\maxNnz & \minNnz-\maxNnz \\
\bottomrule
\end{tabular}
\end{table}

Since both collections consist of ternary schemes, the table also shows the ternary part of the dataset separately: it alone realizes \countZT{} rank patterns, more than the number of schemes in either collection.

The schemes of both collections were not used during the search, but all $\unionPatterns$ rank patterns of these collections are present in the dataset. The remaining schemes of the dataset are inequivalent to every scheme of~\citep{heule2021new, maartensson2026hell}. Together with the collection of Heule et al., whose schemes are pairwise inequivalent, they give at least \numOrbitsLB{} pairwise inequivalent rank-$23$ schemes. The schemes of~\citep{maartensson2026hell} are not added to this bound, since their pairwise inequivalence was not established. For the $\unionPatterns$ patterns shared with the collections, nothing is claimed: the scheme of the dataset may or may not be equivalent to a scheme of some collection with the same pattern, since equivalence was not tested directly.

The dataset and the existing collections are therefore complementary. The collection of Heule et al.\ demonstrates that a single rank pattern may contain many inequivalent schemes, whereas the dataset demonstrates that the number of different rank patterns is far greater than previously observed.

\subsection{The sparsest scheme}
\label{sec:dataset-sparsest}

The sparsest scheme found has $138$ nonzero coefficients and was obtained in June 2026 by GradALS (Section~\ref{sec:als}). To the best of the author's knowledge, no scheme with fewer than $139$ nonzero coefficients was known before that date, neither in the collections of~\citep{heule2021new, maartensson2026hell} nor in the results of earlier searches by other authors. The scheme was not published as a separate work and was only sent by email to several researchers. In September 2026, a post on the KnotTheory.ai blog~\citep{knottheory} described a scheme with $138$ nonzero coefficients found by a support-capped local search over $\mathbb{F}_3$ followed by a lift to the rationals. That scheme uses the wider set of coefficients $\{0\} \cup \{\pm 2^k : -4 \le k \le 4\}$ and contains $30$ non-integer coefficients, whereas the scheme described here contains only $12$. The rank patterns of the two schemes coincide, but their exact equivalence was not verified.

The scheme described here uses the coefficients $\{0, \pm\nicefrac{1}{2}, \pm 1, \pm2\}$, so it belongs to the rational class. It is not part of the dataset: its rank pattern coincides with that of a ternary scheme with $146$ nonzero coefficients, which by the selection rule of Section~\ref{sec:search-procedure} takes its place. The scheme is therefore given separately, in Appendix~\ref{app:sparsest}.

\subsection{Formats and verification}
\label{sec:dataset-formats}

The schemes are provided in three formats: JSON, plain text (\texttt{.txt}) and Maple (\texttt{.m}). The files are named \texttt{3x3x3\_m23\_<hash>\_<class>.<ext>}, where \texttt{<hash>} is the hash of the scheme and \texttt{<class>} is the coefficient class: \texttt{ZT} (ternary), \texttt{Z} (integer) or \texttt{Q} (rational). The JSON file stores the dimension and the rank of the scheme (\texttt{dimension}, \texttt{rank}), the number of nonzero coefficients (\texttt{nnz}), the matrices $U$, $V$, $W$ as lists of rows (\texttt{u}, \texttt{v}, \texttt{w}), with rational coefficients written as strings (\texttt{"-1/2"}), and the invariants $\type$, $f$ and $g$. The text file contains the header line \texttt{3 3 3 23} followed by three lines with the flattened matrices $U$, $V$ and $W$. In the text files of rational schemes, every coefficient is written as a pair of numbers, the numerator and the denominator, so \texttt{-1 2} stands for $-\nicefrac{1}{2}$ and \texttt{1 1} for $1$. The Maple file stores the three matrices $u^{(l)}$, $v^{(l)}$, $w^{(l)}$ of every term as nested lists, with rational coefficients written as \texttt{-1/2}. In all formats, the coefficients of $W$ are stored in the transposed encoding of Section~\ref{sec:preliminaries}.

The correctness of every scheme is checked by verifying the Brent equations~\eqref{eq:brent} in exact arithmetic over integers and rationals. The verification script is included in the repository of the dataset. The same checks, the loading of schemes from all three formats, the computation of the invariants and other utilities are implemented by the \texttt{Scheme} class in the main research repository~\citep{perminov2026fmm}.

%% file: structure/discussion.tex
\section{Discussion}
\label{sec:discussion}

\subsection{Practical value of rank-23 schemes}
\label{sec:discussion-practice}

Rank $23$ is considerably larger than what would be needed to compete with Strassen's algorithm. Applied recursively to blocks, a scheme of rank $r$ for $3 \times 3$ matrices gives the exponent $\log_3 r$, which is smaller than Strassen's $\log_2 7 \approx 2.807$ only if $r < 3^{\log_2 7} \approx 21.85$, that is, for $r \le 21$. Rank $23$ gives $\log_3 23 \approx 2.854$, so none of the schemes of the dataset can improve on Strassen's algorithm in a recursive setting, and in practice they are unlikely to replace it in applications where speed is crucial. The value of the dataset lies elsewhere: it shows how many structurally different schemes exist at the smallest known rank, and it provides a large and diverse starting pool for searches that optimize other properties of schemes.

The most important of these properties is the number of additions, whose record values are listed in the Introduction. The number of nonzero coefficients is only a rough proxy for it, since it ignores the reuse of intermediate sums across linear forms, but it is cheap to compute and serves as a natural first filter. Applying the optimization of additions to the whole dataset, rather than to a handful of schemes, is a natural direction for future work.

\subsection{Lower bounds and the status of rank 23}
\label{sec:discussion-bounds}

Recent lower bounds change the status of rank $23$. The rank of $3 \times 3$ matrix multiplication over $\mathbb{F}_2$ is exactly $23$~\citep{tomar2026rank}, and by reduction modulo two this means that $23$ is optimal for schemes with integer coefficients. The ternary and integer schemes of the dataset are therefore optimal in rank, and the dataset can be viewed as a collection of different optimal schemes. This does not apply to rational schemes with even denominators, such as those with denominator $2$ in the dataset, which cannot be reduced modulo two. Over fields of characteristic zero, the rank is only known to lie between $21$~\citep{Ivanov_A_lower_bound_2026} and $23$, so a scheme of rank $22$ or lower over $\mathbb{Q}$ or $\mathbb{C}$ is not ruled out.

\subsection{Limitations}
\label{sec:discussion-limitations}

The dataset has several limitations, which follow from how it was built and how its schemes are distinguished.

\begin{itemize}
    \item \emph{Completeness and equivalence.} The dataset contains one scheme per rank pattern found by the search. It is neither a complete list of rank patterns nor of orbits. Inequivalence of its schemes is guaranteed by the rank patterns, but equivalence was not tested directly, for example with the algorithm of~\citep{kauers2022normal}, so the dataset gives only a lower bound on the number of orbits and makes no claims about schemes of earlier collections that share a rank pattern with a scheme of the dataset.

    \item \emph{Selection rule.} The coefficient class takes priority over sparsity (Section~\ref{sec:search-procedure}). Hence ternary schemes may exist for patterns represented by integer or rational ones, and sparser schemes with the same pattern may exist; the scheme of Section~\ref{sec:dataset-sparsest} is an example.
\end{itemize}

\subsection{Future directions}
\label{sec:discussion-future}

Several natural continuations follow from the above. The dataset can be used as a source of schemes for optimizing additions, for studying the sparsity and structure of rank-$23$ schemes, and for other properties that depend on the scheme rather than on the rank, such as numerical stability. The same approach, finding schemes with pairwise distinct invariants, can be applied to other small formats. Finally, the search itself can be continued: the dataset keeps growing, and each new rank pattern extends the lower bound on the number of inequivalent schemes. A finer test of equivalence than the rank pattern, such as the algorithm of~\citep{kauers2022normal}, applied to the schemes that share a rank pattern, would turn the lower bound on the number of orbits into a more precise count and would make it possible to compare the dataset with the existing collections scheme by scheme.

%% file: structure/conclusion.tex
\section{Conclusion}
\label{sec:conclusion}

A dataset of \numSchemes{} rank-$23$ schemes for the multiplication of $3 \times 3$ matrices has been presented. Every scheme has its own rank pattern, so all schemes are pairwise inequivalent. All $\unionPatterns$ rank patterns of the collections of Heule et al.~\citep{heule2021new} and M{\aa}rtensson et al.~\citep{maartensson2026hell} are present in the dataset, and all other schemes are inequivalent to every scheme of these collections. Together with the collection of Heule et al., the dataset gives at least \numOrbitsLB{} pairwise inequivalent rank-$23$ schemes. A single rank pattern may thus contain many inequivalent schemes, and, as shown here, the number of different rank patterns is much greater than was previously known.

A scheme with $138$ nonzero coefficients was also found, which is fewer than in any scheme of the publicly available collections known before June 2026. It is given in Appendix~\ref{app:sparsest}. A scheme with the same rank pattern and number of nonzero coefficients was later published in a blog post~\citep{knottheory}.

Since rank $23$ is optimal over $\mathbb{F}_2$ and for schemes with integer coefficients, the dataset can be regarded as a large collection of different optimal schemes, which may serve as a starting pool for optimizing other properties, first of all the number of additions.

%% file: structure/availability.tex
\section{Availability}
\label{sec:availability}

All data and code related to this work are publicly available in the following repositories.
\begin{itemize}
    \item \emph{Dataset}: \url{https://github.com/dronperminov/rank23_3x3_dataset}. The schemes are provided in JSON, plain text and Maple formats, together with a script that verifies the correctness of every scheme.

    \item \emph{Main research repository} on fast matrix multiplication schemes~\citep{perminov2026fmm}: \url{https://github.com/dronperminov/FastMatrixMultiplication}. It contains the \texttt{Scheme} class, which loads schemes from all the formats of the dataset, verifies the Brent equations in exact arithmetic and computes the invariants of Section~\ref{sec:symmetries}, as well as a catalog of schemes and other utilities.

    \item \emph{Flip graph tool}~\citep{perminov2025fast, perminov2026fast}: \url{https://github.com/dronperminov/ternary_flip_graph}. It implements the flip graph search of Section~\ref{sec:search}, including the search over ternary coefficients and over finite fields with Hensel lifting.

    \item \emph{GradALS method}: \url{https://github.com/dronperminov/numerical-fast-matmul}. It implements a numerical search that combines gradient descent with alternating least squares and uses several regularizers to obtain exact schemes with small coefficients (Section~\ref{sec:als}).
\end{itemize}

%% file: structure/acknowledgements.tex
\section*{Acknowledgements}

The author is grateful to Kirill Khoruzhii, Isaac Wood, Erik M{\aa}rtensson, Paul Stankovski Wagner, Joshua Stapleton and Alexandre Sedoglavic for useful conversations and discussions.

%% file: structure/appendix.tex
\section{Rank-23 scheme with 138 nonzero coefficients}
\label{app:sparsest}

The scheme of Section~\ref{sec:dataset-sparsest} is given below in the encoding of Section~\ref{sec:preliminaries}. The matrix $W$ is printed transposed, as $W^T \in \mathbb{Q}^{9\times 23}$, so that its $l$-th column is the vector $w^{(l)}$. The terms are listed in the order produced by the search, not in the sorted form of Section~\ref{sec:polynomial-invariants}. It has $\operatorname{nnz}(U) = 47$, $\operatorname{nnz}(V) = 47$ and $\operatorname{nnz}(W) = 44$. The Brent equations~\eqref{eq:brent} were verified in exact rational arithmetic.

\begin{small}
\setlength{\arraycolsep}{1.8pt}
{\renewcommand{\arraystretch}{1.3}
\[
U = \left[\begin{array}{l *{8}{wr{14pt}}}
   0 & 0 & 0 & 0 & 0 & 0 & 1 & 0 & 0 \\
   0 & 0 & 0 & 0 & 0 & 0 & 0 & 0 & 1 \\
   0 & 0 & 0 & 0 & 0 & 0 & 0 & 1 & 0 \\
   0 & 0 & 0 & 0 & 0 & 0 & 0 & 0 & 1 \\
   0 & 0 & 0 & 0 & 0 & 0 & 1 & 0 & 0 \\
   0 & 0 & 1 & 0 & 0 & 0 & 0 & 0 & 0 \\
   1 & 0 & 2 & 0 & 0 & 0 & 0 & 0 & 0 \\
   0 & 1 & 2 & 0 & 0 & 0 & 0 & 0 & 0 \\
   0 & 0 & 0 & 0 & 0 & 0 & 2 & -1 & 2 \\
   1 & 0 & 0 & 1 & 0 & 0 & -1 & 0 & 0 \\
   0 & 0 & 1 & 0 & 0 & 1 & 0 & 0 & -1 \\
   0 & 1 & 0 & 0 & 1 & 0 & 0 & -1 & 0 \\
   0 & 0 & 0 & 1 & 0 & 0 & 0 & 0 & 0 \\
   0 & 0 & 0 & 0 & 1 & 0 & 0 & 0 & 0 \\
   0 & 0 & 1 & 0 & -\frac{1}{2} & 0 & 0 & 0 & 0 \\
   0 & 0 & 1 & -\frac{1}{2} & 0 & 0 & 0 & 0 & 0 \\
   0 & 1 & 0 & 0 & 1 & 0 & 0 & 0 & 0 \\
   1 & -\frac{1}{2} & 1 & 0 & 0 & 0 & 0 & 0 & 0 \\
   0 & 0 & 0 & 1 & -\frac{1}{2} & 1 & 0 & 0 & 0 \\
   0 & 0 & 0 & 1 & -1 & 0 & 0 & 0 & 2 \\
   0 & 1 & 0 & 1 & 0 & 0 & -2 & 0 & 0 \\
   0 & 0 & 0 & \frac{1}{2} & 0 & 1 & 0 & 0 & -1 \\
   0 & 0 & 0 & 1 & -1 & 0 & 0 & 0 & 0 \\
\end{array}\right]\quad
V = \left[\begin{array}{l *{8}{wr{14pt}}}
   1 & 0 & 0 & 0 & 0 & 0 & 0 & 0 & 0 \\
   0 & 0 & 0 & 0 & 0 & 0 & 1 & 0 & 0 \\
   0 & 0 & 0 & 1 & 0 & 0 & 0 & 0 & 0 \\
   0 & 0 & 0 & 0 & 0 & 2 & 0 & 0 & 1 \\
   0 & 0 & 1 & 0 & 0 & 2 & 0 & 0 & 0 \\
   2 & 0 & 0 & 2 & 0 & 0 & -1 & 0 & 0 \\
   1 & 0 & 0 & 0 & 0 & 0 & 0 & 0 & 0 \\
   0 & 0 & 0 & 1 & 0 & 0 & 0 & 0 & 0 \\
   0 & 0 & 0 & 0 & 0 & 1 & 0 & 0 & 0 \\
   1 & -1 & -1 & 0 & 0 & 0 & 0 & 0 & 0 \\
   0 & 0 & 0 & 0 & 0 & 0 & 1 & -1 & -1 \\
   0 & 0 & 0 & 1 & -1 & -1 & 0 & 0 & 0 \\
   0 & 0 & 1 & 0 & 0 & 1 & 0 & 0 & -\frac{1}{2} \\
   0 & 1 & 0 & 0 & 1 & 0 & 0 & -\frac{1}{2} & 0 \\
   0 & 1 & 0 & 2 & 0 & 0 & 0 & -1 & 0 \\
   2 & -1 & 0 & 0 & 0 & 0 & 0 & 0 & -1 \\
   0 & \frac{1}{2} & 0 & -1 & 1 & 0 & 0 & 0 & 0 \\
   0 & 1 & 0 & 0 & 0 & 0 & 0 & 0 & 0 \\
   0 & 0 & 0 & 0 & 0 & 0 & 0 & 1 & 0 \\
   0 & 0 & 0 & 0 & 0 & -1 & 0 & \frac{1}{2} & 0 \\
   0 & \frac{1}{2} & 0 & 0 & 0 & -1 & 0 & 0 & 0 \\
   0 & 0 & 0 & 0 & 0 & 0 & 0 & 1 & 1 \\
   0 & \frac{1}{2} & 0 & 0 & 0 & 0 & 0 & -\frac{1}{2} & 0 \\
\end{array}\right]
\]
\[
W^T = \left[\begin{array}{*{23}{wr{14pt}}}
   0 & 0 & 0 & 0 & 0 & -1 & 1 & 1 & 0 & 0 & 0 & 0 & 0 & 0 & 0 & 0 & 0 & 0 & 0 & 0 & 0 & 0 & 0 \\
   0 & 1 & 0 & 0 & 0 & 1 & 0 & 0 & 0 & 0 & 1 & 0 & 0 & 0 & -1 & -1 & 0 & 0 & 0 & 0 & 0 & 1 & 1 \\
   1 & 1 & 1 & 0 & 0 & 0 & 0 & 0 & 0 & 0 & 0 & 0 & 0 & 0 & 0 & 0 & 0 & 0 & 0 & 0 & 0 & 0 & 0 \\
   0 & 0 & 0 & 0 & 0 & 0 & 0 & 1 & 0 & 0 & 0 & 0 & 0 & -1 & -1 & 0 & 1 & 1 & 0 & 0 & 0 & 0 & 0 \\
   0 & 0 & 0 & 0 & 0 & 0 & 0 & 0 & 0 & 0 & 0 & 0 & 0 & 1 & 0 & 0 & 0 & 0 & 1 & 0 & 0 & 0 & 2 \\
   0 & 0 & 1 & 0 & 0 & 0 & 0 & 0 & 1 & 0 & 0 & 1 & 0 & 0 & 0 & 0 & 1 & 0 & 0 & 1 & -1 & 0 & 1 \\
   -1 & 0 & 0 & 0 & 1 & 0 & 1 & 0 & 0 & -1 & 0 & 0 & -1 & 0 & 0 & -1 & 0 & -1 & 0 & 0 & -1 & 0 & 0 \\
   0 & 0 & 0 & 1 & 0 & 0 & 0 & 0 & 0 & 0 & 0 & 0 & 1 & 0 & 0 & 0 & 0 & 0 & -1 & 1 & 0 & 1 & 0 \\
   0 & 0 & 0 & 1 & 1 & 0 & 0 & 0 & -1 & 0 & 0 & 0 & 0 & 0 & 0 & 0 & 0 & 0 & 0 & 0 & 0 & 0 & 0 \\
\end{array}\right]
\]}
\end{small}

\medskip
\noindent\textbf{Invariants.} The invariants of Section~\ref{sec:symmetries} for this scheme are

\begin{itemize}
    \item $\type(x,y,z) = 4x^2y^2z^2 + x^2yz + xy^2z + 2xyz^2 + 15xyz$;
    \item $f(x,y,z) = 24x^2y^2z^2 + 8x^2yz + 8xy^2z + 8xyz^2 + 90xyz$;
    \item $g(\omega) = \omega^{29} + 2\omega^{28}$.
\end{itemize}

In terms of the rank pattern of Section~\ref{sec:rank-pattern}, the scheme consists of $15$ terms in which all three matrices $u^{(l)}$, $v^{(l)}$, $w^{(l)}$ have rank~$1$, four terms in which all three have rank~$2$, and four terms in which exactly one of the three matrices has rank~$2$.

%% file: references.bib
@article{strassen1969gaussian,
  title={Gaussian elimination is not optimal},
  author={Strassen, Volker},
  journal={Numerische mathematik},
  volume={13},
  number={4},
  pages={354--356},
  year={1969},
  publisher={Springer}
}

@techreport{brent1970algorithms,
  title={Algorithms for matrix multiplication},
  author={Brent, Richard P},
  year={1970},
  institution={Stanford University, Stanford, CA, USA}
}

@article{laderman1976noncommutative,
  title={A noncommutative algorithm for multiplying $3\times 3$ matrices using 23 multiplications},
  journal={Bulletin of the American Mathematical Society},
  author={Laderman, Julian D},
  year={1976}
}

@article{fawzi2022discovering,
  title={Discovering faster matrix multiplication algorithms with reinforcement learning},
  author={Fawzi, Alhussein and Balog, Matej and Huang, Aja and Hubert, Thomas and Romera-Paredes, Bernardino and Barekatain, Mohammadamin and Novikov, Alexander and R. Ruiz, Francisco J and Schrittwieser, Julian and Swirszcz, Grzegorz and others},
  journal={Nature},
  volume={610},
  number={7930},
  pages={47--53},
  year={2022},
  publisher={Nature Publishing Group UK London}
}

@article{deza2023fast,
  title={Fast matrix multiplication without tears: A constraint programming approach},
  author={Deza, Arnaud and Liu, Chang and Vaezipoor, Pashootan and Khalil, Elias B},
  journal={arXiv preprint arXiv:2306.01097},
  year={2023}
}

@article{novikov2025alphaevolve,
  title={AlphaEvolve: A coding agent for scientific and algorithmic discovery},
  author={Novikov, Alexander and V{\~u}, Ng{\^a}n and Eisenberger, Marvin and Dupont, Emilien and Huang, Po-Sen and Wagner, Adam Zsolt and Shirobokov, Sergey and Kozlovskii, Borislav and Ruiz, Francisco JR and Mehrabian, Abbas and others},
  journal={arXiv preprint arXiv:2506.13131},
  year={2025}
}

@inproceedings{kauers2023flip,
  title={Flip graphs for matrix multiplication},
  author={Kauers, Manuel and Moosbauer, Jakob},
  booktitle={Proceedings of the 2023 International Symposium on Symbolic and Algebraic Computation},
  pages={381--388},
  year={2023}
}

@inproceedings{arai2024adaptive,
  title={Adaptive flip graph algorithm for matrix multiplication},
  author={Arai, Yamato and Ichikawa, Yuma and Hukushima, Koji},
  booktitle={Proceedings of the 2024 International Symposium on Symbolic and Algebraic Computation},
  pages={292--298},
  year={2024}
}

@inproceedings{moosbauer2025flip,
  title={Flip graphs with symmetry and new matrix multiplication schemes},
  author={Moosbauer, Jakob and Poole, Michael},
  booktitle={Proceedings of the 2025 International Symposium on Symbolic and Algebraic Computation},
  pages={233--239},
  year={2025}
}

@inproceedings{khoruzhii2026faster,
  title={Faster algorithms for structured matrix multiplication via flip graph search},
  author={Khoruzhii, Kirill and Gel{\ss}, Patrick and Pokutta, Sebastian},
  booktitle={Proceedings of the 2026 International Symposium on Symbolic and Algebraic Computation},
  pages={227--236},
  year={2026}
}

@article{kauers2025exploring,
  title={Exploring the meta flip graph for matrix multiplication},
  author={Kauers, Manuel and Wood, Isaac},
  journal={arXiv preprint arXiv:2510.19787},
  year={2025}
}

@article{perminov2026fast,
  title={Fast Matrix Multiplication in Small Formats: Discovering New Schemes with an Open-Source Flip Graph Framework},
  author={Perminov, Andrew},
  journal={arXiv preprint arXiv:2603.02398},
  year={2026}
}

@article{perminov2025fast,
  title={Fast Matrix Multiplication via Ternary Meta Flip Graphs},
  author={Perminov, Andrew},
  journal={arXiv preprint arXiv:2511.20317},
  year={2025}
}

@article{sedoglavic2025yet,
  title={Yet another catalogue of fast matrix multiplication algorithms, 2019},
  author={Sedoglavic, Alexandre},
  journal={URL: http://fmm. univ-lille. fr, accessed: november},
  year={2026}
}

@article{lacelle2026catalog,
  title={A catalog of fast matrix multiplication algorithms with frontier-closure search},
  author={Lacelle, Benoit Chatain},
  journal={arXiv preprint arXiv:2606.13408},
  year={2026}
}

@article{stapleton2025sixty,
  title={A 60-addition, rank-23 scheme for exact $3\times 3$ matrix multiplication},
  author={Stapleton, Joshua},
  journal={arXiv preprint arXiv:2508.03857},
  year={2025}
}

@article{martensson2026fiftynine,
  title={A Rank 23 Algorithm for Multiplying $3\times 3$ Matrices with an Arithmetic Complexity of 59},
  author={M{\aa}rtensson, Erik and Wagner, Paul Stankovski and Stapleton, Joshua},
  journal={arXiv preprint arXiv:2601.05272},
  year={2025}
}

@article{perminov2025addition,
  title={A 58-Addition, Rank-23 Scheme for General 3x3 Matrix Multiplication},
  author={Perminov, Andrew},
  journal={arXiv preprint arXiv:2512.21980},
  year={2025}
}

@article{sun2026fiftysix,
  title={An Exact 56-Addition, Rank-23 Scheme for General $3\times 3$ Matrix Multiplication},
  author={Sun, Yinqi},
  journal={arXiv preprint arXiv:2604.27645},
  year={2026}
}

@article{karunaratne2026fiftyfive,
  title={55 Additions Suffice for $3\times 3$ Matrix Multiplication at Rank 23},
  author={Karunaratne, Samurdhi and Idamekorala, Anushka},
  journal={arXiv preprint arXiv:2607.28676},
  year={2026}
}

@article{sidebottom2026fiftyfive,
  title={A 55-Addition Rank-23 Scheme for $3\times 3$ Matrix Multiplication via Exact Two-Sided Minimization},
  author={Sidebottom, Greg},
  journal={arXiv preprint arXiv:2610.27451},
  year={2026}
}

@article{stapleton2026alternative,
  title={A 51-Addition Alternative-Basis Kernel for Rank-23 $3\times 3$ Matrix Multiplication},
  author={Stapleton, Joshua and Perminov, Andrew},
  journal={arXiv preprint arXiv:2609.32388},
  year={2026}
}

@article{heule2021new,
  title={New ways to multiply $3\times 3$-matrices},
  author={Heule, Marijn JH and Kauers, Manuel and Seidl, Martina},
  journal={Journal of Symbolic Computation},
  volume={104},
  pages={899--916},
  year={2021},
  publisher={Elsevier}
}

@article{johnson1986noncommutative,
  title={Noncommutative Bilinear Algorithms for $3\times 3$ Matrix Multiplication},
  author={Johnson, Rodney W and McLoughlin, Aileen M},
  journal={SIAM Journal on Computing},
  volume={15},
  number={2},
  pages={595--603},
  year={1986},
  publisher={SIAM}
}

@article{winograd1971multiplication,
  title={On multiplication of $2\times 2$ matrices},
  author={Shmuel Winograd},
  journal={Linear Algebra and its Applications},
  year={1971},
  volume={4},
  pages={381-388},
  url={https://api.semanticscholar.org/CorpusID:123566199}
}

@article{hopcroft1971minimizing,
  title={On minimizing the number of multiplications necessary for matrix multiplication},
  author={Hopcroft, John E and Kerr, Leslie R},
  journal={SIAM Journal on Applied Mathematics},
  volume={20},
  number={1},
  pages={30--36},
  year={1971},
  publisher={SIAM}
}

@misc{Ivanov_A_lower_bound_2026,
  author = {Ivanov, Andrey},
  doi = {10.5281/zenodo.23324839},
  license = {Apache-2.0},
  month = oct,
  title = {{A lower bound of 21 for $3 \times 3$ matrix multiplication over every field}},
  url = {https://github.com/AIV5/mmrank21},
  version = {0.1.1},
  year = {2026}
}

@article{tomar2026rank,
  title={The rank of $3\times 3$ matrix multiplication over $\mathbb{F}_2$ is 23},
  author={Tomar, Tejasvi Singh},
  journal={arXiv preprint arXiv:2610.11823},
  year={2026}
}

@inproceedings{martensson2024beast,
  title={The number of the beast: Reducing additions in fast matrix multiplication algorithms for dimensions up to 666},
  author={M{\aa}rtensson, Erik and Wagner, Paul Stankovski},
  booktitle={2025 Proceedings of the Conference on Applied and Computational Discrete Algorithms (ACDA)},
  pages={47--60},
  year={2025},
  organization={SIAM}
}

@article{degroote1978varieties,
  title={On Varieties of Optimal Algorithms for the Computation of Bilinear Mappings},
  author={de Groot, HF},
  journal={Theoretical Computer Science},
  volume={7},
  number={2},
  pages={127--148},
  year={1978}
}

@inproceedings{hopcroft1973duality,
  title={Duality applied to the complexity of matrix multiplications and other bilinear forms},
  author={Hopcroft, John and Musinski, Jean},
  booktitle={Proceedings of the fifth annual ACM symposium on Theory of computing},
  pages={73--87},
  year={1973}
}

@article{burichenko2014symmetries,
  title={On symmetries of the Strassen algorithm},
  author={Burichenko, Vladimir P},
  journal={arXiv preprint arXiv:1408.6273},
  year={2014}
}

@misc{knottheory,
  author       = {Sharma, Chandragupt},
  title        = {Chasing zeros: a rank-23 scheme for {$3\times3$} matrix multiplication with support 138},
  howpublished = {KnotTheory.ai Research Notes (blog post)},
  year         = {2026},
  month        = sep,
  url          = {https://knottheory.ai/blogs/matrix-multiplication-3x3-support-138},
  note         = {Published September 30, 2026. Accessed November 5, 2026}
}

@article{maartensson2026hell,
  title={Hell’s Bells: A Neural Network Pipeline for Ternary Fast Matrix Multiplication Algorithms},
  author={M{\aa}rtensson, Erik and Wagner, Paul Stankovski and Stapleton, Joshua},
  journal={Cryptology ePrint Archive},
  url = {https://eprint.iacr.org/2026/1688},
  year={2026}
}

@misc{perminov2026fmm,
  author       = {{Perminov, Andrew}},
  title        = {Fast Matrix Multiplication research repository},
  year         = {2026},
  publisher    = {GitHub},
  journal      = {GitHub repository},
  howpublished = {\url{https://github.com/dronperminov/FastMatrixMultiplication}},
  note         = {Accessed November 5, 2026}
}

@misc{heule2019matrix,
  author       = {Heule, Marijn J. and Kauers, Manuel and Seidl, Martina},
  title        = {Matrix Multiplication Repository},
  year         = {2019},
  howpublished = {\url{http://www.algebra.uni-linz.ac.at/research/matrix-multiplication/}}
}

@article{nielsen202652,
  title={A 52-Addition, Rank-23 Scheme for Exact 3 x 3 Matrix Multiplication},
  author={Nielsen, Hugo},
  journal={arXiv preprint arXiv:2609.06588},
  year={2026}
}

@inproceedings{kauers2022normal,
  title={A normal form for matrix multiplication schemes},
  author={Kauers, Manuel and Moosbauer, Jakob},
  booktitle={International Conference on Algebraic Informatics},
  pages={149--160},
  year={2022},
  organization={Springer}
}

@software{fmm_add_reduction,
author = {Stankovski Wagner, Paul},
license = {CC-BY-NC-SA-4.0},
month = may,
title = {{fmm\_add\_reduction}},
url = {https://github.com/werekorren/fmm_add_reduction},
year = {2025}
}
